\documentclass[twoside,twocolumn,showpacs,floatfix]{revtex4}
\usepackage{amssymb,amsmath}
\usepackage[pdftex]{graphicx}
\usepackage{color}

\begin{document}

\title{Traveling Cluster Pairs in a System of Phase Oscillators with Positive and Negative Couplings under a Periodic Driving Field}

\author{Jungzae \surname{Choi}}
\affiliation{Department of Physics and Department of Chemical Engineering, Keimyung University, Daegu 42601, Korea}
\author{M.Y. \surname{Choi}}
\affiliation{Department of Physics and Astronomy and Center for Theoretical Physics, Seoul National University, Seoul 08826, Korea}
\author{Moon Sung \surname{Chung} and Byung-Gook \surname{Yoon}}
\thanks{Author to whom correspondence should be addressed. E-mail: bgyoon@ulsan.ac.kr}
\affiliation{Department of Physics, University of Ulsan, Ulsan 44610, Korea}

\begin{abstract}
We investigate numerically the clustering behavior of a system of phase oscillators with positive and negative couplings under a periodic external driving field with a bimodal distribution of driving phases. The phase distribution and the mean speed of the traveling state, as well as the order parameter for synchronization, are computed as the driving amplitude is varied. We observe that the periodically-driven system can also host traveling states for parameters in the same range as those for the case of a system without a driving field.
The traveling speed is found to depend non-monotonically on the driving amplitude.
In particular, oscillators divide into four clusters and move in pairs. Further, depending on the driving amplitude, two kinds of traveling mode arise: pairs of clusters traveling in the same direction (symmetric mode) and in opposite directions (antisymmetric mode). In the latter case (antisymmetric traveling mode), the average phase speed of the whole system apparently vanishes. A phenomenological argument for such behavior is given.
\end{abstract}

\pacs{05.45.Xt}
\keywords{Coupled oscillators, Periodic driving field, Bimodal driving distribution, Phase distribution, Cluster}
\maketitle

\section{Introduction}

A system of sinusoidally-coupled phase oscillators, first introduced by Winfree~\cite{ref:Winfree} and later refined by Kuramoto~\cite{ref:Kuramoto0,ref:Kuramoto}, serves as a typical model for collective synchronization. When the coupling strength between oscillators is increased sufficiently, the system becomes ordered, forming a cluster of oscillators that move together on a phase circle.
Though there have appeared many variations of the Kuramoto model~\cite{ref:extension},
the following extension of the model attracts interest with regard to the formation of clusters and their motion: When both repulsive and attractive couplings are appropriately introduced in the model~\cite{ref:hs1}, two clusters may be formed with the phase split by a difference of less than $\pi$ radians. These two clusters, maintaining the phase difference, travel on the phase circle; this is dubbed a traveling state. Recently, the dynamics of the traveling
state has been studied in terms of the phase speed of the clusters~\cite{ref:ccy14-2}.
According to that study, the positive (attractively coupled) cluster chases after the negative (repulsively coupled) cluster indefinitely.

Other extensions of the model include a periodic driving field, which may give rise to periodic synchronization: The order parameter may oscillate if the coupling is strong enough~\cite{ref:psynch}.
When the driving strength is large enough, a synchronized state is usually
followed by a desynchronized state in a single period, and the duration of the former depends on the parameters of the system.
Recently, we have shown that two clusters may be formed when a system of globally-coupled oscillators is driven periodically, which can explain the periodic
synchronization-desynchronization behavior~\cite{ref:ccy14}.

In this work, we consider a model that incorporates both features of the two types of extension:
a system of positively and negatively coupled oscillators under a periodic sinusoidal driving 
field with a bimodal distribution of driving phases.
% In a pair one cluster travels with the other in the same direction (symmetric %mode) or in the opposite direction (antisymmetric mode).
We observe, via numerical calculations, that four clusters emerge and travel in paired motion.
In particular, two modes of traveling exist, depending on the phase of the driving field:
The two pairs travel either in the same direction (symmetric mode) or in opposite directions
(antisymmetric mode).
In the latter mode, albeit the parameters are the same, the system appears not to be in a traveling state.
%We give some phenomenological understanding on these behaviors.

This paper consists of four sections: Section II describes the
oscillator model and its dynamics. In Section III, numerical results
and the phenomenology of the two types of traveling mode are presented. Finally, a brief
summary is given in Section IV.

\section{Numerical calculation}

We consider a system of $N$ oscillators, the $i$th of which has
an intrinsic frequency $\omega_i$. An oscillator is described by
its phase $\phi_i$ and is coupled globally to the other oscillators.
The dynamics of such a coupled oscillator system is governed
by the set of Langevin equations of motion ($i=1,...,N$):
\begin{equation} \label{model}
 \dot{\phi_i} =\omega_i -  \frac{K_i}{N}\sum_{j=1}^N  \sin(\phi_i-\phi_j)+ I_{0} \cos(\Omega t+\alpha_i) ,
\end{equation}
where the intrinsic frequencies $\{\omega_i\}$ are assumed to be symmetrically distributed according to the Lorentzian distribution $g(\omega)=(\gamma/\pi) (\omega^2 + \gamma^2 )^{-1}$.
The second term on the right-hand side represents
sinusoidal interactions between oscillators with strength $K_i$,
which is positive or negative depending on
whether or not the oscillator follows the mean field $\Delta$ in Eq.~(\ref{eqn}) below.
Specifically, it is taken from the distribution
$\Gamma(K)=(1-p)\delta(K{-}K_n)+p\delta(K{-}K_p)$, where
$K_n \,(<0)$ and $K_p \,(>0)$ are negative and positive coupling
constants, with $p$ denoting the probability that an oscillator
has a positive coupling constant. The final term describes
a driving field with period $\tau = 2\pi/\Omega$,
where the driving phases, the $\alpha_{i}$'s, are distributed
according to the bimodal distribution $f(\alpha)=(1/2)[\delta(\alpha )+\delta(\alpha{-}\pi )]$,
independently of the natural frequencies $\omega_i$.
In other words, the driving term consists of $I_0\cos \Omega t$ and $I_0\cos \Omega (t+\tau/2)$, time shifted by half the period $\tau/2 = \pi/\Omega$, with equal probabilities.

In order to measure the synchronization of the system, we
introduce the complex {\em order parameter}
\begin{equation} \label{deforder}
 \Psi \equiv \frac{1}{N} \sum_{j=1}^N e^{i \phi_j}
       = \Delta e^{i\theta} ,
\end{equation}
where the magnitude $\Delta$ measures the synchronization of the oscillators
with the average phase $\theta$. The order parameter defined in
Eq.~(\ref{deforder}) allows us to reduce Eq.~(\ref{model}) to a
{\em single} decoupled equation:
\begin{equation} \label{eqn}
 \dot{\phi_i} = \omega_i - K_i \Delta \sin(\phi_i -\theta) + I_{0} \cos(\Omega t+\alpha_i).
\end{equation}

In this work we resort to numerical methods to investigate the system.
%Using the second-order Runge-Kutta-Helfand-Greenside algorithm, we integrate
Using the second-order Runge-Kutta algorithm, we integrate
Eq.~(\ref{eqn}) with the time step $\delta t=0.01$ for the system size
$N=2000$. Initially ($t=0$), the $\phi_i$'s are %either set to be zero or
taken to be randomly distributed between $0$ and $2\pi$ for all $i$. We fix the period
%of driving to be $\tau=5.12$ and the positive coupling constant $K_p =1$.
of the driving field to be $\tau=5.12$, with the positive coupling constant $K_p =1$ and
$p=0.6$. Most of the results in this work remain qualitatively unchanged
even when we choose other values of $p$ for which a traveling state emerges. %%%% to meet the refree's question (i) and (iii)

After the initial transient behavior,
the system reaches stationarity and we obtain the time series of the
order parameter as well as the time evolution of the phase distribution.
Here, we examine the populations in a finite number of regions in phase space.
Specifically, we divide one cycle of the phase into
72 different regions, the $k$th of which is defined by the phase
interval $(k\pi/36{-}\pi/72, \,k\pi/36{+}\pi/72]$ (modulo $2\pi$), and we obtain
the number $n_k$ of oscillators belonging to the $k$th region
($k =0,...,71$).
We also obtain the mean traveling speed $\langle w \rangle$ averaged
over 30 ensembles, with each undergoing its own time evolution.
Here, the traveling speed $w$ is obtained in the following way:
At each time, we calculate the average value of the phase velocities $w_i=\dot{\phi}_i$ and take
the time average over $10$ periods of the driving field. Then, we take the absolute value to get $w$.

\section{Results and discussion}

\begin{figure}
%\centerline{\epsfig{file=Dvst1.eps,width=8.5cm}}
\includegraphics[width=8.5cm]{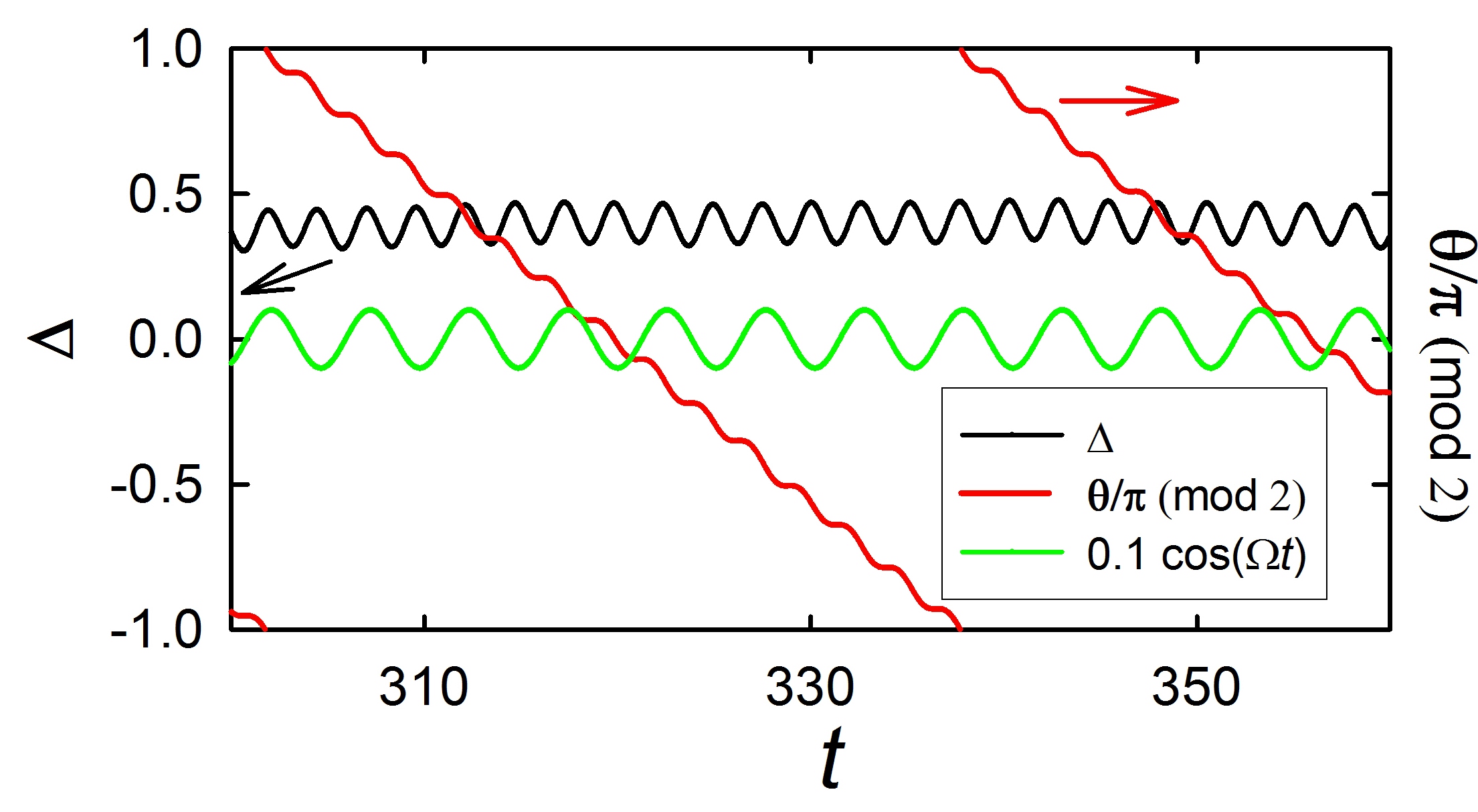}
\caption{(color online) Magnitude $\Delta$ and phase in units of $\pi$, $\theta/\pi$ (modulo 2) of the order parameter versus time $t$ for a system of $N=2000$ oscillators with $I_{0}=1$, $p=0.6$, $K_n =-0.5$, and $\gamma=0.01$.
Also plotted is the cosine function of the driving field.
}
\label{fig:Dt_1}
\end{figure}

\begin{figure}
%\centerline{\epsfig{file=dis1.eps,width=8.5cm}}
\includegraphics[width=8.5cm]{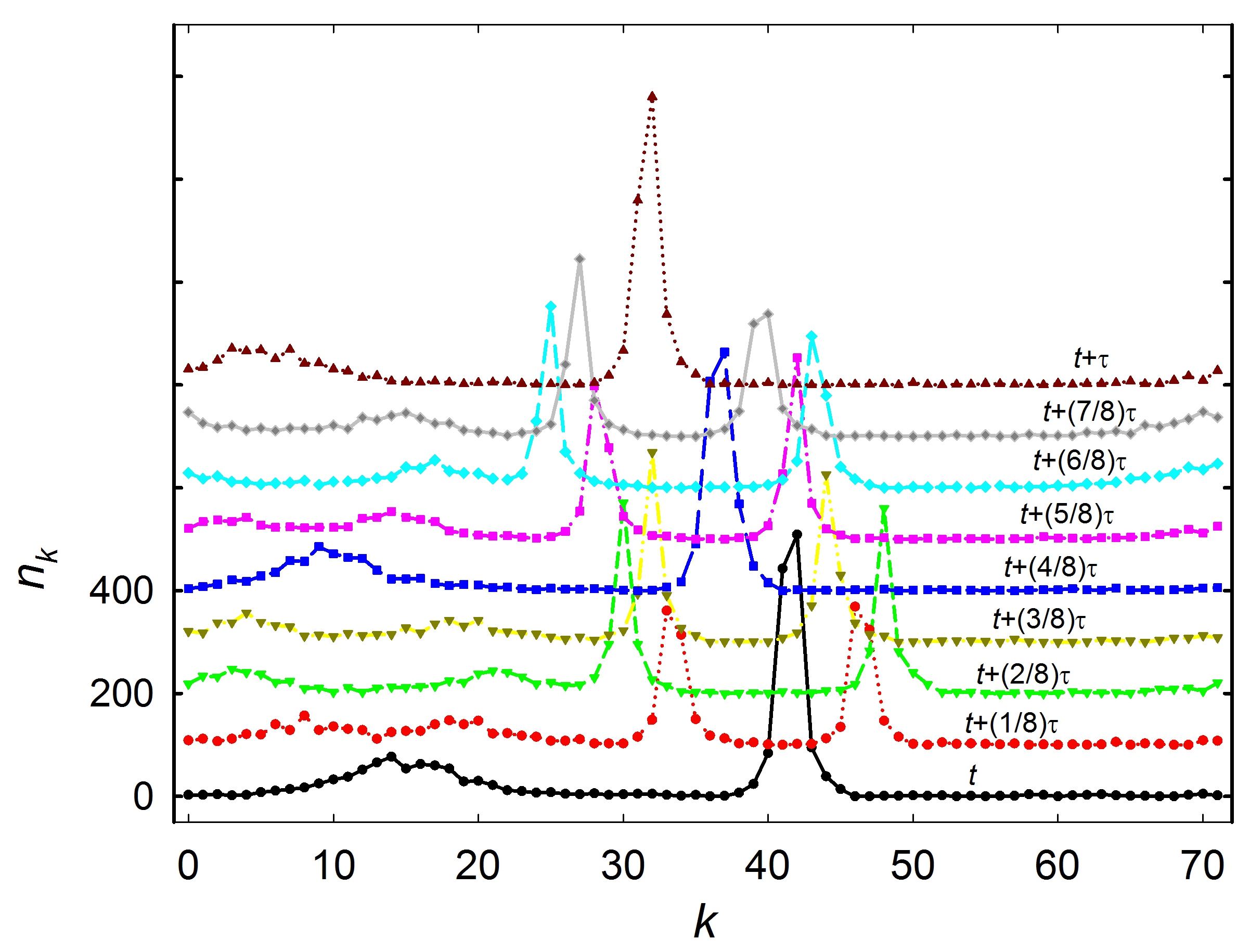}
\caption{(color online) Number $n_k$ of oscillators belonging to region $k$
with phase $\phi$ satisfying $k\pi/36{-}\pi/72 < \phi \leq k\pi/36{+}\pi/72$ (modulo $2\pi$)
at nine successive times during one period in the same system as in Fig.~\ref{fig:Dt_1}.
From bottom to top, the nine data sets describe the distribution of oscillators $n_k$ at times $t+(n/8)\tau$ for $n=0, 1, \ldots, 8$, respectively.
The lines between data points are merely guides to the eye.
For clarity, the curves have been successively shifted upward by 100.
%The lowermost curve is for the earliest time $t$ and the uppermost
%one is for the time after one period of driving from $t$.
}
\label{fig:dis1}
\end{figure}

Figure~\ref{fig:Dt_1} present the time evolution of the magnitude $\Delta$ and the phase in units of $\pi$, $\theta/\pi$ (modulo 2) of the order parameter in a system of $N=2000$ oscillators for $I_{0}=1$, $p=0.6$, $K_n =-0.5$, and $\gamma=0.01$.
Also plotted is the cosine function of the driving field.
Without a driving field, as described in Ref.~5,
the traveling state emerges when the magnitude of the negative coupling constant
is less than unity ($|K_n /K_p |<1$), and nearly half the oscillators are positive ($p \approx 0.5$).
For small driving amplitudes, for instance $I_0 =1$, the system exhibits periodic synchronization, as revealed by the behavior of $\Delta$ in Fig.~\ref{fig:Dt_1}. What is different from the traveling state without a driving field is the appearance of ripples both in $\Delta$ and in $\theta$, reflecting the periodic driving field in the system. Presented in Fig.~\ref{fig:dis1} is the distribution of oscillators, i.e., the number $n_k$ of oscillators belonging to the $k$-th region, at nine successive times in one period of the driving field from $t$ (lowermost) to $t+\tau$ (uppermost).
%Similar to the case of no periodic driving there arises the traveling wave state~\cite{ref:hs1},
Here, clearly, the two clusters, the larger one consisting of positive oscillators
and the smaller of negative ones,
%emerge and rotate in the phase space. %when the value of cosine function in Eq.~(\ref{model}) is small.
emerge twice in one period of the driving field.
Subsequently, each cluster divides into two clusters in accordance with the driving phase $\alpha$.
% one moving in the opposite direction to the other.
In other words, four clusters characterized by the sign of coupling and by the driving phase
(see below) emerge. After half the period of the driving field, two positive/negative clusters merge and generate a traveling peak.

%\begin{figure}
%\centerline{\epsfig{file=Dvst2.eps,width=8.5cm}}
%\includegraphics[width=8.5cm]{Dvst2.pdf}
%\caption{(color online) Magnitude, $\Delta$, and phase over $\pi$, $\theta/\pi$ (modulo 2), of order parameter
%vs time $t$ for a system with $N=2000$, $I_{0}=2$, $p=0.6$, $K_n =-0.5$ and $\gamma=0.01$.
%Also drawn is the cosine function of the driving.
%}
%\label{fig:Dt_2}
%\end{figure}

\begin{figure}
%\centerline{\epsfig{file=Dvst1-6.eps,width=8.5cm}}
\includegraphics[width=8.5cm]{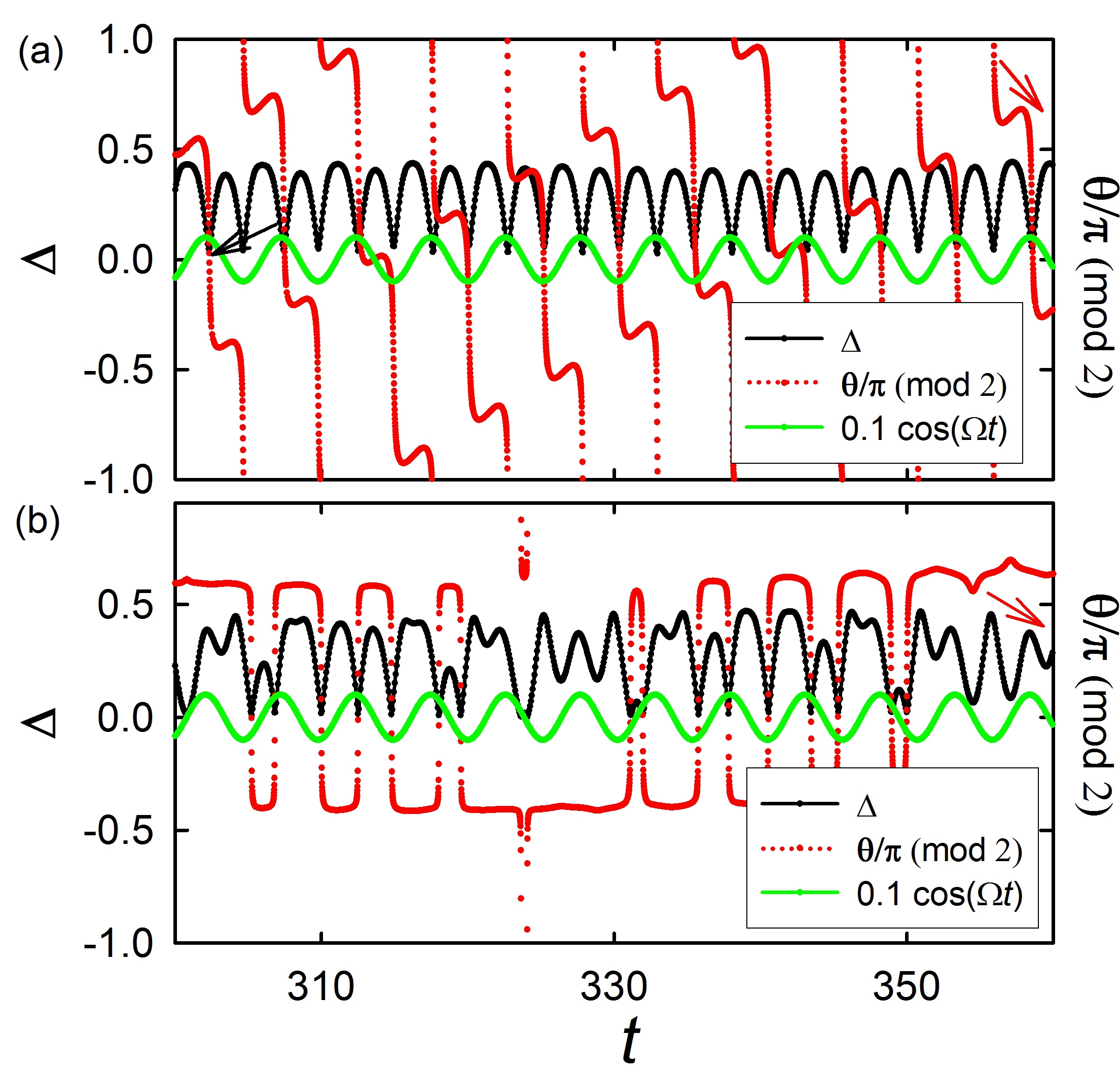}
\caption{ (color online) Magnitude $\Delta$ and phase $\theta$ in units of $\pi$ (modulo 2)
of the order parameter versus time $t$ for a system of $N=2000$ oscillators with $I_{0}=1.6$, $p=0.6$, $K_n =-0.5$, and $\gamma=0.01$.
Also plotted is the cosine function of the driving field.
Data sets in (a) and (b) have been obtained from the same system, but starting from different initial configurations.
}
\label{fig:Dt1.6}
\end{figure}

As the driving strength is increased, the cluster behavior of the system becomes more complex:
%Each cluster divides into two clusters in accordance with the sign of the driving amplitude $I_0$.
% one moving in the opposite direction to the other.
%In other words, there emerge four clusters characterized by the sign of coupling and the driving amplitude
%(see below).
Two clusters move in a pair in the same direction with a phase split while two pairs travel
in the same direction or in opposite directions for some values of $I_0$,
depending on the initial configuration.
% in pairs, After the half period, two divided clusters become overlapped again with the merged peaks
%which have been traveled.
%Remarkably, for some values of $I_0$, the clusters do not appear to travel at all as presented
%in Fig.~\ref{fig:Dt1.6},
%Thus, for some values of $I_0$, the clusters for some systems do not appear to travel at all,
Thus, in some cases, the clusters do not appear to travel at all, as illustrated in Fig.~\ref{fig:Dt1.6},
which presents the time evolution of $\Delta$ and $\theta/\pi$ (modulo 2) in a system of $N=2000$ oscillators with $I_0=1.6$, $p=0.6$, $K_n=-0.5$, and $\gamma=0.01$ for two different initial configurations of phases. Compared with the behavior in Fig.~\ref{fig:Dt_1}, the oscillatory behavior of $\Delta$ is more pronounced, and the phase changes more rapidly in Fig.~\ref{fig:Dt1.6}(a). However, the mean traveling speed, which may also be estimated from the slope of a line joining the local maxima of $\theta$, is apparently smaller than that in Fig.~\ref{fig:Dt_1}.

\begin{figure}
%\centerline{\epsfig{file=dis1-6n.eps,width=8.5cm}}
\includegraphics[width=8.5cm]{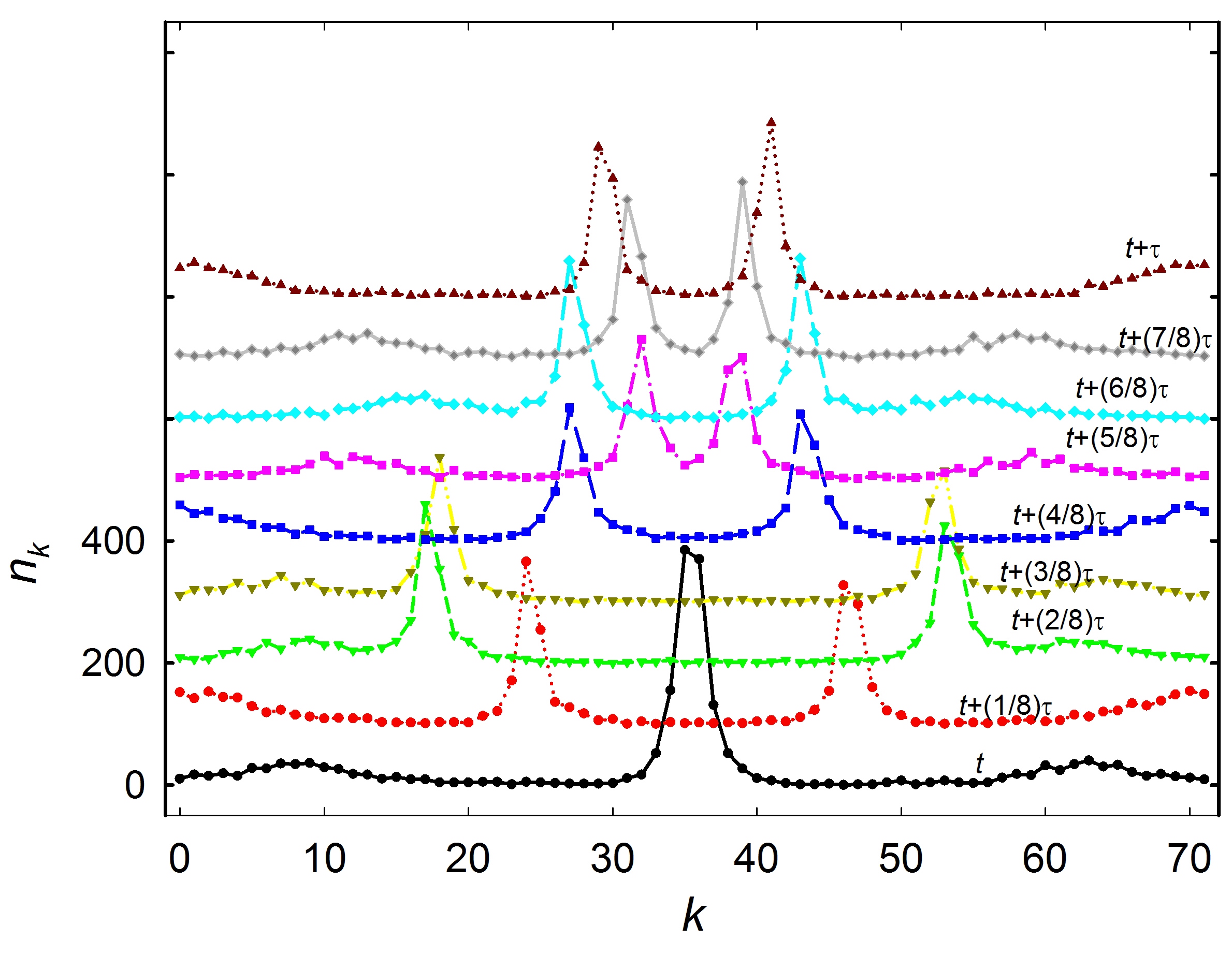}
\caption{ (color online) Number $n_k$ of oscillators belonging to region $k$ at nine successive times,
as labeled, with the data obtained from the same system as that in Fig.~\ref{fig:Dt1.6}(b).
The lines between data points are merely guides to the eye.
For clarity, the curves have been successively shifted upward by 100.
}
\label{fig:dis1.6n}
\end{figure}

\begin{figure}
\includegraphics[width=8.5cm]{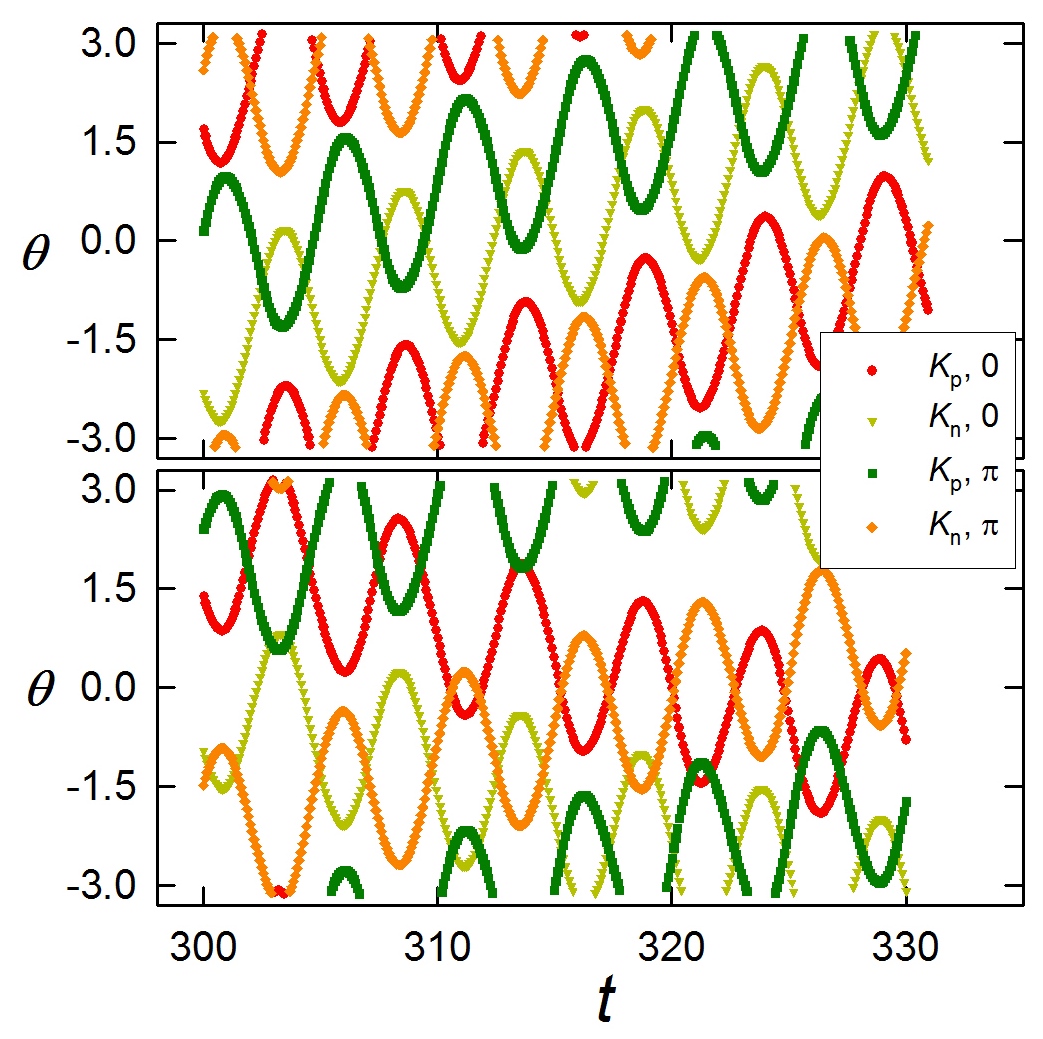}
\caption{ (color online) Average phase $\theta$ of each cluster versus time $t$, as determined
from the order parameters of each cluster in the systems corresponding to Figs.~\ref{fig:Dt1.6}(a) and (b), respectively.
The indices $K$ and $\alpha$ labeling clusters are given in the legend.
}
\label{fig:Tht1.6n}
\end{figure}

On the other hand, Fig.~\ref{fig:Dt1.6}(b) reveals that the behavior of the same system can be significantly different for a different initial configuration:
The order parameter $\Delta$ is not periodic in time and the mean traveling speed is apparently zero,
as the phase $\theta$ remains almost constant except for sporadic and abrupt changes of sign.
To confirm the apparently aperiodic behavior of $\Delta$,
we display in Fig.~\ref{fig:dis1.6n} the phase distribution $n_k$ corresponding to the data of Fig.~\ref{fig:Dt1.6}(b).
Manifestly, the phase distribution does not vary periodically in time:
A single cluster at time $t$ splits into two clusters, which do not merge into a single one, but remain
%as two at time $t+\tau$. Moreover, the two pairs of clusters move in opposite directions.
as two at time $t+\tau$. Moreover, the two pairs of clusters travel in opposite directions.
This is made clearer by probing the time evolution of the average phases of the aforementioned clusters,
which are calculated from the order parameters for all clusters in the system in Fig.~\ref{fig:Dt1.6} and are displayed in Fig.~\ref{fig:Tht1.6n}.
The average phase of each cluster is labeled according to the signs of the coupling ($K=K_p$ or $K_n$)
and the driving phase ($\alpha =0$ or $\pi$). While in Fig.~\ref{fig:Tht1.6n}(a) the two pairs of clusters consisting of positive and negative oscillators with the driving phase $\alpha =0$ and with the phase $\alpha =\pi$ travel in the same direction, the two pairs of clusters travel in opposite directions in Fig.~\ref{fig:Tht1.6n}(b).
Note that positive and negative oscillators that belong to the same pair (the same driving phase $\alpha$)
move in the same direction. We may, thus, conclude that two modes of traveling are possible in a driven oscillator system: the symmetric mode (one-way motion) and the antisymmetric mode (two-way motion).
Regardless of the traveling mode, both pairs do travel, and the phase split is always less than $\pi$.
The average phase velocity, however, vanishes for the antisymmetric mode because two pairs travel in opposite directions at the same traveling speed.

%This traveling is caused by the fact that the separation of
%two clusters are less than $\pi$ radian in phase space. In this case, the coupling
%term in Eq.~(\ref{model}) make a non-zero contribution to traveling speed $w$, while%
%the first and the driving term do not.

%When the negative coupling is introduced in a system of which the
%$\langle \Delta \rangle$ is near the local minima in Fig.~\ref{fig:wI0},
%non-traveling state as well as traveling state occurs.

\begin{figure}
%\centerline{\epsfig{file=wvsI0.eps,width=8.5cm}}
\includegraphics[width=8.5cm]{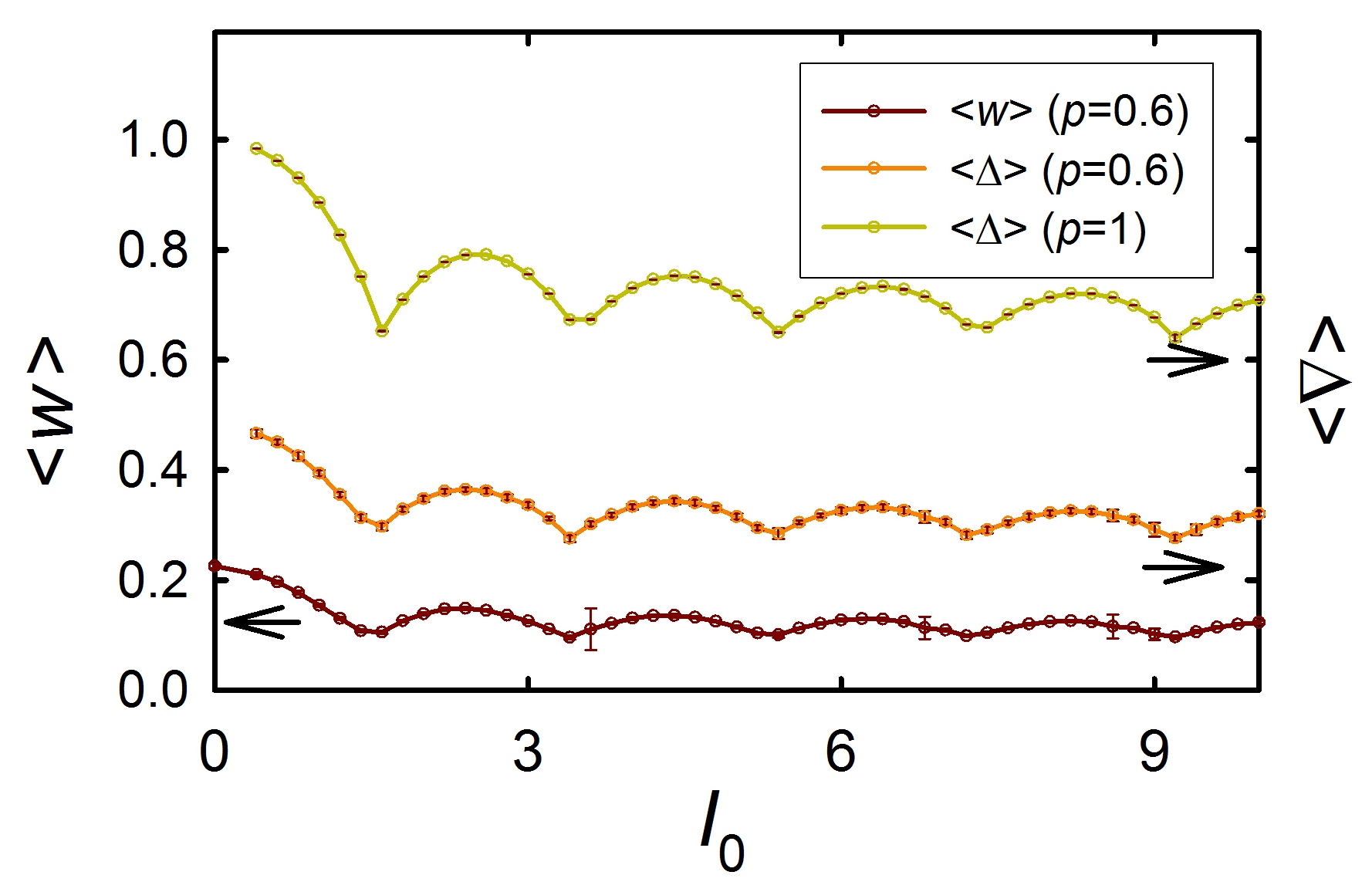}
\caption{(color online) Mean traveling speed $\langle w \rangle$ and average
order parameter $\langle \Delta \rangle$ versus driving amplitude $I_{0}$
for a system of $N=2000$ oscillators with $p=0.6$, $K_n =-0.5$, and $\gamma=0.01$.
Also shown is $\langle \Delta \rangle$ versus $I_{0}$ for a system
without negative coupling ($p=1$).
Error bars denote standard deviations, and lines are merely guides to the eye.
}
\label{fig:wI0}
\end{figure}

Whenever the antisymmetric mode comes into play,
the mean traveling speed remains relatively small
%only for values of $I_0$ %at which
as shown in Fig.~\ref{fig:wI0}, which presents the mean traveling speed
$\langle w \rangle$ and the average order parameter $\langle \Delta \rangle$ versus $I_{0}$
in a system of $N=2000$ oscillators with $p=0.6$, $K_n =-0.5$, and $\gamma=0.01$.
Here, data only for the symmetric mode have been used in the calculation of ensemble averages.
Also shown is $\langle \Delta \rangle$ versus $I_{0}$ for the system with only positive couplings ($p=1$).
We observe that $\langle \Delta \rangle$ has local minima and exhibits an oscillatory behavior.
% The oscillatory behavior of $\langle \Delta \rangle$ also occurs in systems with only positive couplings as well.
Note that the data shown in Fig.~\ref{fig:Dt1.6} describe the case $I_{0}=1.6$,
which corresponds to the first local minimum in Fig.~\ref{fig:wI0}.
%, as is shown in the same figure: the data presenting
%$\langle \Delta \rangle$ vs $I_{0}$ for systems with $N=2000$, $p=1$ and $\gamma=0.01$.

\begin{figure}
%\centerline{\epsfig{file=Dvstp1.eps,width=8.5cm}}
\includegraphics[width=8.5cm]{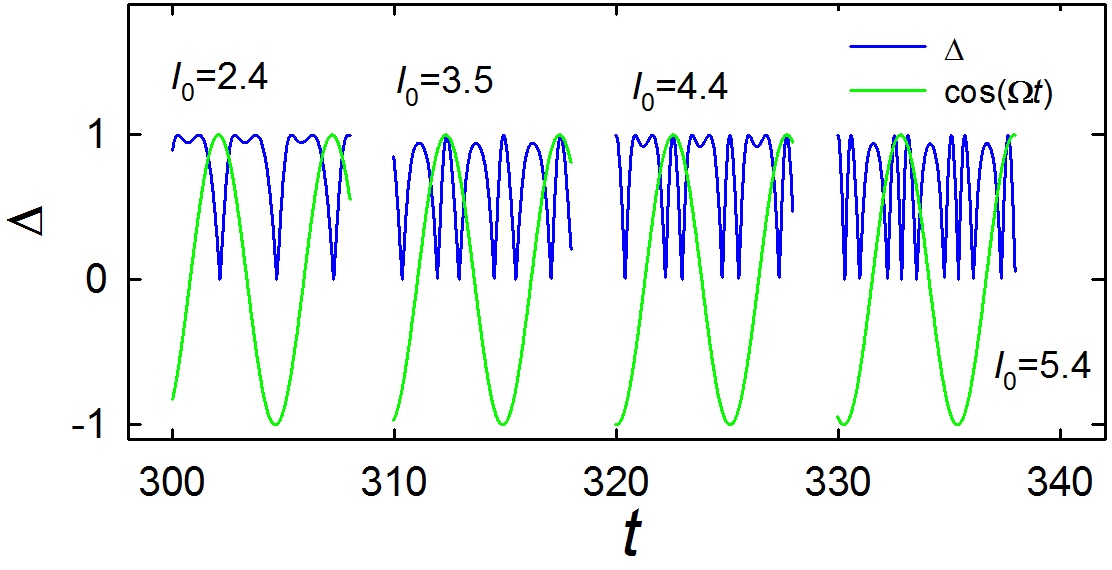}
\caption{(color online) Magnitude $\Delta$ of the order parameter versus time $t$ for a system
of $N=2000$ positive oscillators ($p =1$) with $\gamma=0.01$ and $I_{0} = 2.4, 3.5, 4.4,$ and $5.4$.
The cosine function of the driving is also plotted. The second ($I_0 =3.5$) and fourth ($I_0 =5.4$) curves
correspond to the values of $I_{0}$ at which the order parameter reaches local minima in Fig.~\ref{fig:wI0}.
%Note that there is no negative coupling in these systems.
}
\label{fig:Dt_p1}
\end{figure}

In order to understand this behavior, we consider a system of positive oscillators, i.e., oscillators interacting via only positive couplings. Figure~\ref{fig:Dt_p1} shows how the order parameter $\Delta$ evolves
in time $t$ for four different values of $I_{0}$, two of which ($I_0=3.5$ and $5.4$) correspond to
the local minima of $\Delta$ in Fig.~\ref{fig:wI0} while the other two ($I_0=2.4$ and $4.4$) correpond to the local maxima.
%Also drawn is the cosine function of the driving.
For the latter two values of $I_{0}$, as indicated by the maxima of $\Delta$, the two clusters overlap or are
located rather close to each other. This occurs when the driving term is small, i.e., the cosine function of the driving field takes a small value.
The order parameter $\Delta$ reaches its maximum before the cosine function changes its sign.
The once overlapping clusters become separated again as time goes on.
When the direction of the driving field is reversed, the two clusters overlap again and then separate.
On the other hand, for the former values of $I_{0}\, (=3.5$ and $5.4)$, $\Delta$ barely reaches a maximum
as the sign of the cosine function changes; % near the center of broader peaks.
therefore, the values of $\langle \Delta \rangle$ are smaller than those of the latter cases
that exhibit double peaks in the time evolution. % has the larger value of $\langle\Delta\rangle$.

%\begin{figure}
%\includegraphics[width=8.5cm]{fig8.jpg}
%\caption{(color online) Mean traveling speed $\langle w \rangle$ versus %driving amplitude $I_{0}$
%for three values of $p$ as shown in the legend.
%The other parameters are the same as those in Fig.~\ref{fig:wI0}.
%Error bars denote standard deviations, and lines are merely guides to the %eye.
%}
%\label{fig:wI0_p}
%\end{figure}

We now return to systems consisting of both positive and negative oscillators. %%%following: for revision
Suppose that pairs of clusters are established in the antisymmetric mode by the initial conditions
or during the transient time. Two positive (or negative) clusters travel in opposite directions.
If the driving amplitude $I_{0}$ takes one of those values corresponding to the local maxima
in Fig.~\ref{fig:wI0}, these clusters will have sufficient time to interact and to get to the symmetric mode.
Recall the presence of double peaks of $\Delta $ in the system of positive oscillators
shown in Fig.~\ref{fig:Dt_p1}, signifying the existence of two clusters for some time.
Because the clusters are not much separated during this time,
the interactions between clusters via coupling begins to have an effect,
so that the motion of the pairs becomes symmetric after several periods of the driving field.
%%%%%%%%%%%%%%
When $I_0$ takes a value near one of the local minima in Fig.~\ref{fig:wI0},
on the other hand, this is less likely because the two pairs of clusters do not have enough time to evolve
into the symmetric mode. As a result, the system may remain in the antisymmetric mode. The symmetric and the anti-symmetric modes are expected to appear (almost) periodically with $I_0$ because the oscillators tend to follow the driving field; if the system is in the (anti-)symmetric mode at some value of $I_0$, then the system is also in the (anti-)symmetric mode at the other value of $I_0$ increased by $\Delta I$ (in our case $\Delta I\approx 2$), except that the oscillators move longer on a phase circle.
%Meanwhile we observe that no antisymmetric mode takes place for local %maximum values of $I_0$, unlike the case of local minimum values of $I_0$ %where a considerable fraction of the system indeed exhibits the %antisymmetric mode. At those values of $I_0$, the traveling speed is low %even for the symmetric mode; the order parameter is small and clusters are %more broadly distributed. In particular, the broad distribution of negative %oscillators is expected to reduce the traveling speed of clusters.
%Such effects of negative clusters on reducing the traveling speed can be %observed in Fig.~\ref{fig:wI0_p},
%which displays the mean traveling speed $\langle w \rangle$ versus the %driving amplitude $I_{0}$
%for three values of $p$ shown in the legend. As the fraction of negative %oscillators is increased or $p$ is reduced, the overall traveling speed %becomes lower.
The oscillatory behavior persists for smaller values of $p$ with lower values of $\langle w \rangle$ (data not shown).
%In this manner, we have the oscillating behavior of the phase velocity in Fig.~\ref{fig:wI0}. %%% revision
%%%%%%%%%%%%%%%%%

Finally, the aperiodicity appearing in Fig.~\ref{fig:Dt1.6}(b) can be understood as follows:
While clusters in the symmetric mode are rather far apart in phase from each other, in the antisymmetric mode, the distance between two clusters often becomes small, which makes the interactions between these clusters considerable. Such effects accelerate or decelerate the motions of clusters, resulting in the aperiodicity.

\section{Summary}

We have numerically studied systems of oscillators interacting via positive and negative couplings under a periodic sinusoidal driving field with the driving phases following a bimodal distribution.
Observed is the formation of two pairs of clusters, each consisting of positive and negative clusters
according to the phase of the driving field. Depending on the amplitude of the driving field, two pairs of clusters travel in the same direction or in opposite directions during each period of the driving field, which we call the symmetric and the antisymmetric modes of traveling, respectively.
We have also found that the mean traveling speed of the symmetric mode depends non-monotonically on the
driving amplitude and that the traveling speed is small in the antisymmetric mode.
Discussions on how the symmetric and the antisymmetric modes arise are given.

\acknowledgments
This work was supported in part by the 2014 Research Fund of the University of Ulsan.

\end{document}